\documentclass[iop]{emulateapj}
\usepackage{graphicx,apjfonts}

\shorttitle{\maxi\ in quiescence}
\shortauthors{Homan et al.}

\newcommand{\maxi}{MAXI J1659--152}
\newcommand{\cha}{{\it Chandra}}
\newcommand{\ledd}{$L_{\rm Edd}$}
\newcommand{\xmm}{{\it XMM-Newton}}
\newcommand{\sw}{{\it Swift}}
\newcommand{\porb}{$P_{\rm orb}$}

\newcommand{\pbif}{$P_{\rm bif}$}

\newcommand{\flux}{erg\,s$^{-1}$\,cm$^{-2}$}
\newcommand{\lum}{erg\,s$^{-1}$}

\begin{document}

\title{The X-ray properties of the black hole transient MAXI J1659--152 in quiescence} 
\author{Jeroen Homan\altaffilmark{1},  Joel K.\ Fridriksson\altaffilmark{2}, Peter G.\ Jonker\altaffilmark{3}, David M.\ Russell\altaffilmark{4,5}, Elena Gallo\altaffilmark{6}, Erik Kuulkers\altaffilmark{7}, Nanda Rea\altaffilmark{8}, and Diego Altamirano\altaffilmark{2}} 
\altaffiltext{1}{Kavli Institute for Astrophysics and Space Research, Massachusetts Institute of Technology, 70 Vassar Street, Cambridge, MA 02139, USA; jeroen@space.mit.edu}
\altaffiltext{2}{Astronomical Institute ``Anton Pannekoek", University of Amsterdam, Postbus 94249, 1090 GE Amsterdam, The Netherlands}
\altaffiltext{3}{SRON, Netherlands Institute for Space Research, Sorbonnelaan 2, 3584 CA Utrecht, The Netherlands}
\altaffiltext{4}{ Instituto de Astrof\'isica de Canarias (IAC), E-38200 La Laguna, Tenerife, Spain}
\altaffiltext{5}{Departamento de Astrof\'isica, Universidad de La Laguna (ULL), E-38206 La Laguna, Tenerife, Spain}
\altaffiltext{6}{Department of Astronomy, University of Michigan, 500 Church Street, Ann Arbor, MI 48109, USA}
\altaffiltext{7}{European Space Astronomy Centre (ESA/ESAC), Science Operations Department, E-28691 Villanueva de la Ca\~nada (Madrid), Spain}
\altaffiltext{8}{Institute of Space Sciences (CSIC-IEEC), Campus UAB, Faculty of Science, Torre C5-parell, E-08193 Barcelona, Spain}

\begin{abstract}

We present new {\it Chandra} X-ray observations of the transient black hole X-ray binary \maxi\ in quiescence. These observations were made more than one year after the end of the source's 2010--2011 outburst. We detect the source at a 0.5--10 keV flux of $2.8(8)\times10^{-15}$ \flux, which corresponds to a luminosity of $\sim$$1.2\times10^{31}$ $(\frac{d}{{\rm 6\,kpc}})^2$ \lum. This level, while being the lowest at which the source has been detected, is within factors of $\sim$2 of the levels seen at the end of the initial decay of the outburst and soon after a major reflare of the source. The quiescent luminosity of \maxi, which is the shortest-orbital-period black hole X-ray binary ($\sim$2.4 hr), is lower than that of neutron-star X-ray binaries with similar periods. However, it is higher than the quiescent luminosities found for black hole X-ray binaries with orbital periods $\sim$2--4 times longer. This could imply that a minimum quiescent luminosity may exist for black hole X-ray binaries, around orbital periods of $\sim$5--10 hr, { as predicted by binary-evolution models for the mass transfer rate}.  Compared to the hard state we see a clear softening of the power-law spectrum in quiescence, from an index of 1.55(4) to an index of 2.5(4). We constrain the luminosity range in which this softening starts to $(0.18$--$6.2)\times10^{-5}$ ($\frac{d}{{\rm 6\,kpc}})^2$ ($\frac{M}{8\,M_\odot}$) \ledd, which is consistent with the ranges inferred for { other sources}.

\end{abstract}

\keywords{accretion, accretion disks -- X-rays: binaries -- X-rays: individual (\maxi)}

\section{Introduction}\label{sec:intro}

Low-mass X-ray binaries (LMXBs) are systems in which a low-mass donor star transfers mass via Roche-lobe overflow onto a neutron star (NS) or black hole (BH). Many of the LMXBs are transient systems. These transients undergo occasional outbursts during which they typically reach peak luminosities of $\sim$0.01--1 times the Eddington luminosity (\ledd), but they spend most of their time in quiescence, with much lower luminosities. { Here we adopt the definition of quiescence (for BH LMXBs) by \citet{plgajo2013}, $l_x = L_{\rm 0.5-10\,keV}/L_{\rm Edd}<10^{-5}$. We note that finding a source at $l_x<10^{-5}$ does not necessarily mean that a source is detected at its minimum quiescent luminosity, as sources have been found with $l_x$  as low as a few times $10^{-9}$ \citep{gamcna2001}. } The nature of the accretion flow in quiescence is still a matter of debate. Proposed explanations for the very low quiescent luminosities include  radiatively inefficient flows \citep{nayi1994} and low net accretion rates in the inner regions as the result of disk winds \citep{blbe1999} or jets \citep{fegajo2003}. 

\begin{figure}[t] 
\centerline{\includegraphics[width=8cm]{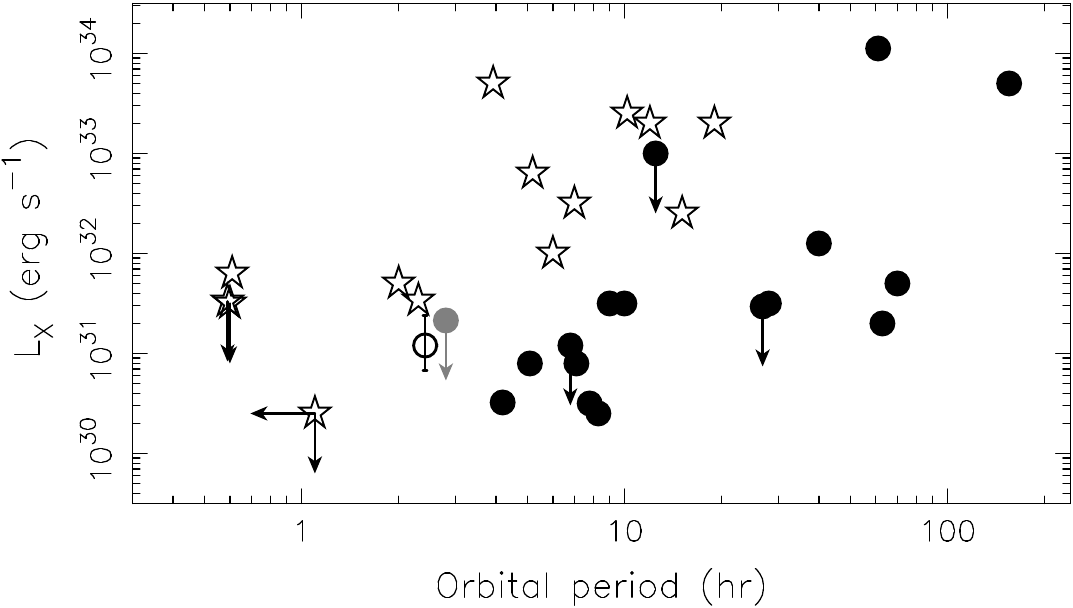}}
\caption{Quiescent 0.5--10 keV luminosities of NS (stars) and BH (circles) X-ray binaries, as a function of the orbital period. The luminosity of \maxi\ (based on observation 7/8) is shown as an open circle, for an assumed distance of 6 kpc \citep{jomiho2012}. The error bar on the data point for \maxi\ reflects the uncertainty in the distance to the source. The gray circle represents Swift J1357.2$-$0933.
Arrows indicate upper limits on luminosity or orbital period.  Based on data presented in \citet{gahojo2008}, \citet{rejone2011}, and \citet{remi2011}.} 
\label{fig:orbital}
\end{figure}

In the last decade, the high { sensitivities} of \cha\ and \xmm\  { have  opened up the possibility of}  detailed X-ray studies of quiescent LMXBs down to Eddington ratios as low as $\sim$$10^{-8}$ \ledd\ \citep{gamcna2001,habala2003}. These observations have revealed that, when comparing LMXBs with similar orbital periods (\porb), quiescent NS systems have on average higher luminosities than quiescent BH systems, by factors of $\sim$10--100 \citep{gamcna2001}. This can clearly be seen in Figure \ref{fig:orbital}. It is important to consider systems with the same \porb;  at a given \porb\ BH and NS LMXBs are assumed to have similar quiescent mass accretion rates \citep{meesna1999}. 
Expected mass transfer rates for quiescent BHs and NSs were calculated by \citet{meesna1999}, and they showed that for quiescent LMXBs there should exist a minimum mass transfer rate that stems directly from the existence of a bifurcation
orbital period, \pbif. Below this period the mass transfer is driven by
gravitational wave radiation and above it, it is
dominated by the nuclear evolution of the secondary star. Specifically, the mass transfer rate 
increases with decreasing orbital period below \pbif, while 
it increases with \porb\ above \pbif. 
For a wide range of donor masses, { the results of \citet{meesna1999} imply}  $P_{\rm bif}\sim$5--10 hr for BHs and $P_{\rm bif}\sim$3--5 hr for NSs.

The luminosity difference between quiescent BH and NS LMXBs has been interpreted as evidence for the presence of an event horizon in BH LMXBs \citep{gamcna2001}, although it may also be the result of jet-dominated states in quiescent BH LMXBs \citep{fegajo2003}. Interestingly, the lowest-luminosity quiescent source currently known ($L_{\rm X}<2.4\times10^{30}$ erg\,s$^{-1}$) is an NS LMXB, 1H 1905+000 \citep{jostch2007}. However, this source is likely an ultra-compact with an orbital period less than 80 minutes \citep{jobane2006}, and therefore falls in a \porb\ range ($<$4 hr) in which, until recently, no quiescent BH LMXBs had been observed.  

Another result of the  \cha\ and \xmm\ observations of quiescent LMXBs concerns the nature of the quiescent accretion flow in BH systems.The quiescent state of  BH LMXBs has often been considered a low-luminosity extension of the so-called low-hard state. Indeed, in terms of radio/X-ray flux correlations, which likely trace the evolution of the accretion (in)flow and/or jet outflow, quiescent BH LMXBs appear to follow the main relation seen in the low-hard state \citep{gafemi2006,gamife2012}. However, recent observations suggest that in terms of X-ray spectral shape considerable evolution occurs in the accretion flow as some sources approach quiescence \citep{cotoka2006}, with spectral power-law photon indices that are steeper ($\sim$2.2--2.5) than those seen in the low-hard state ($\sim$1.5).  { A recent study of the quiescent spectra of ten BH LMXBs by \citet{plgajo2013} suggests that, once these sources reach $l_x\sim10^{-5}$, their spectra saturate at power-law indices of $\sim2.08\pm0.07$. Based on their findings \citet{plgajo2013} argue that quiescence does not appear to represent a distinct spectral state separated from the low-hard state.} Several mechanisms could be responsible for the observed steepening toward quiescence, such as, e.g., a non-linear dependence of mass accretion rate on the inner-disk radius, as expected in the presence of outflows \citep[see discussion in][]{cotoka2006},  changes in the  properties of a Comptonizing corona \citep{tokaka2004,sopado2011}, { or the jet's cooling break shifting through  the X-ray band \citep{plgajo2013}}.

In this paper we present \cha\ observations of \maxi, an X-ray transient that was discovered in 2010 September  with the {\it Swift} Burst Alert Telescope \citep{mahoma2010}. Although it was originally thought to be a gamma-ray burst, optical spectra obtained with the Very Large Telescope/X-shooter \citep{deflwi2010} and X-ray observations with {\it RXTE} \citep{kayaal2010} strongly suggested that \maxi\ is an LMXB with a BH primary. During an outburst in 2010/2011 that lasted more than nine months, the source was extensively observed with various X-ray \citep{kahoal2011,mumost2011,yaalka2012}, optical/near-infrared \citep{rulebe2010,kakael2012} and radio observatories \citep{vakoka2011,mimajo2011,pavabe2013}. These observations revealed that the source made several state transitions and showed behavior similar to that seen in many other transient BH LMXBs. The distance to \maxi\ has been estimated using various methods, with a most likely range of 4.5--8.5 kpc \citep{keroma2011,kakael2012,ko2012,kukobe2013,jomiho2012}. Following \citet{jomiho2012}, we adopt a distance of 6 kpc in this paper.

{\it XMM-Newton} and {\it Swift} observations made during the rise and maximum of the outburst revealed the presence of dips in the X-ray light curves, from which an orbital period of 2.414$\pm$0.005 hr was derived \citep{kukobe2013,keroma2011}. This makes \maxi\ the BH LMXB with the shortest known orbital period. Such a short orbital period is of particular interest to one of the issues described earlier: the difference between NS and BH quiescent luminosities at low \porb.

\maxi\ was already observed with \cha\ during the initial  decay of its outburst, a subsequent three-month reflare, and soon after the source appeared to have reached quiescence. The results of these observations were reported by \citet{jomiho2012}. { They found the source in a quiescent state during most of their observations.  The minimum quiescent luminosity of \maxi\ was determined from two observations taken shortly after the reflare; it} falls at the high end of what is expected for its orbital period (assuming a distance of 6 kpc), although it is still fainter than quiescent NS LMXBs with similar \porb\ values. One explanation for this higher than expected luminosity could be that at the time of the last \cha\ observations the source still had not reached its { minimum} quiescent luminosity. The new \cha\ observations presented here were made about a year after the source had entered quiescence. { They allow us to test whether the quiescent luminosity reported by \citet{jomiho2012} was close to a minimum luminosity, or if the source had declined even further. We also present a more detailed study of the spectral softening in \maxi.}

\section{Observations and data analysis}

\maxi\ has been observed  eight times with {\it Chandra}. A log of the observations can be found in Table \ref{tab:obs}. All observations were made with the back-illuminated S3 CCD chip of the Advanced CCD Imaging Spectrometer \citep[ACIS;][]{gabafo2003}.  The first six observations were already analyzed by \citet{jomiho2012}. Two new observations, made in 2012 July, were added to the analysis presented in this work. Observations 6--8 were made in VFAINT mode, which allows for  better background cleaning, while the others were made in FAINT mode. All observations were analyzed using CIAO 4.4, CALDB 4.5.5.1, and ACIS Extract version 2012nov1 \citep{brtofe2010}. As a first step, the {\tt chandra$\_$repro} script was run to reprocess  the data from all the observations. The data were checked for episodes of enhanced background, but none were found. Images in the 0.5--7.0 keV band were extracted for each observation to search for extended emission.  Further analysis was performed with the help of ACIS Extract. 

\begin{table*}
\footnotesize
\caption{A Log of {\it Chandra} Observations of \maxi\ and Spectral Fit Results}
\label{tab:obs}
\begin{center}
\begin{tabular}{cccccccc}
\hline
Obs.\ No. & Obs-ID & Start Date/Time  & Exposure  & Net Count Rate & $\Gamma$ & Flux$^a$ & Goodness \\
& & (UT) & (ks) & 0.3--7 keV (counts s$^{-1}$) & & (erg\,s$^{-1}$\,cm$^{-2}$) &  (\%) \\
\hline\hline
1 & 12438 & 2011 Apr 14 23:05:18 & 6.4 & $3.7(2)\times10^{-2}$      & 1.87(12)      & $4.5(4)\times10^{-13}$ & 0.1 \\
2 & 12439 & 2011 Apr 23 17:59:11 &  9.1 & $8.7(9)\times10^{-3}$     & 1.8(2)          & $1.1(2)\times10^{-13}$ & 0.9 \\
3 & 12440 & 2011 May 03 07:09:45 & 13.6 & $4.3(2)\times10^{-4}$   & 2.5(4)$^{\rm b}$  & $7(3)\times10^{-15}$ & 2.5$^{\rm b}$\\
4 & 12441 & 2011 May 12 05:03:10 & 18.1 & $6.67(6)\times10^{-1}$ & 1.55(4)$^{\rm c}$ & $1.38(8)\times10^{-11}$ & 34\\
5 & 12442 & 2011 Aug 15 19:59:16 & 30.8 & $2.9(11)\times10^{-4}$ & 2.5(4)$^{\rm b}$   & $3.4(12)\times10^{-15}$ & 2.5$^{\rm b}$\\
6 & 12443 & 2011 Oct 12 12:45:46 & 90.7 & $4.3(7)\times10^{-4}$    & 2.5(4)$^{\rm b}$ & $5.0(8)\times10^{-15}$ & 2.5$^{\rm b}$ \\
7 & 14454 & 2012 Jul 03 00:52:46 & 39.4 & $1.1(6)\times10^{-4}$     & 2.5(4)$^{\rm b}$  & $1.7(9)\times10^{-15}$ & 2.5$^{\rm b}$\\
8 & 13731 & 2012 Jul 07 02:50:23 & 35.5 & $2.2(8)\times10^{-4}$     & 2.5(4)$^{\rm b}$  & $3.7(12)\times10^{-15}$ & 2.5$^{\rm b}$\\
7+8 & & & & & &  $2.8(8)\times10^{-15}$ & \\
\hline

\end{tabular}	
\end{center}

{\bf Notes.} Errors on the fit parameters reflect the 1$\sigma$ uncertainties. \\
      $^{\rm a}$Unabsorbed flux in the  0.5--10 keV band.\\
$^{\rm b}$These observations were fitted together with power-law indices linked. \\
$^{\rm c}$Pile-up model parameter alpha is 0.11 $\pm$ 0.10.
\vspace{0.2cm}

\end{table*}

\begin{figure}[b] 
\centerline{\includegraphics[width=8cm]{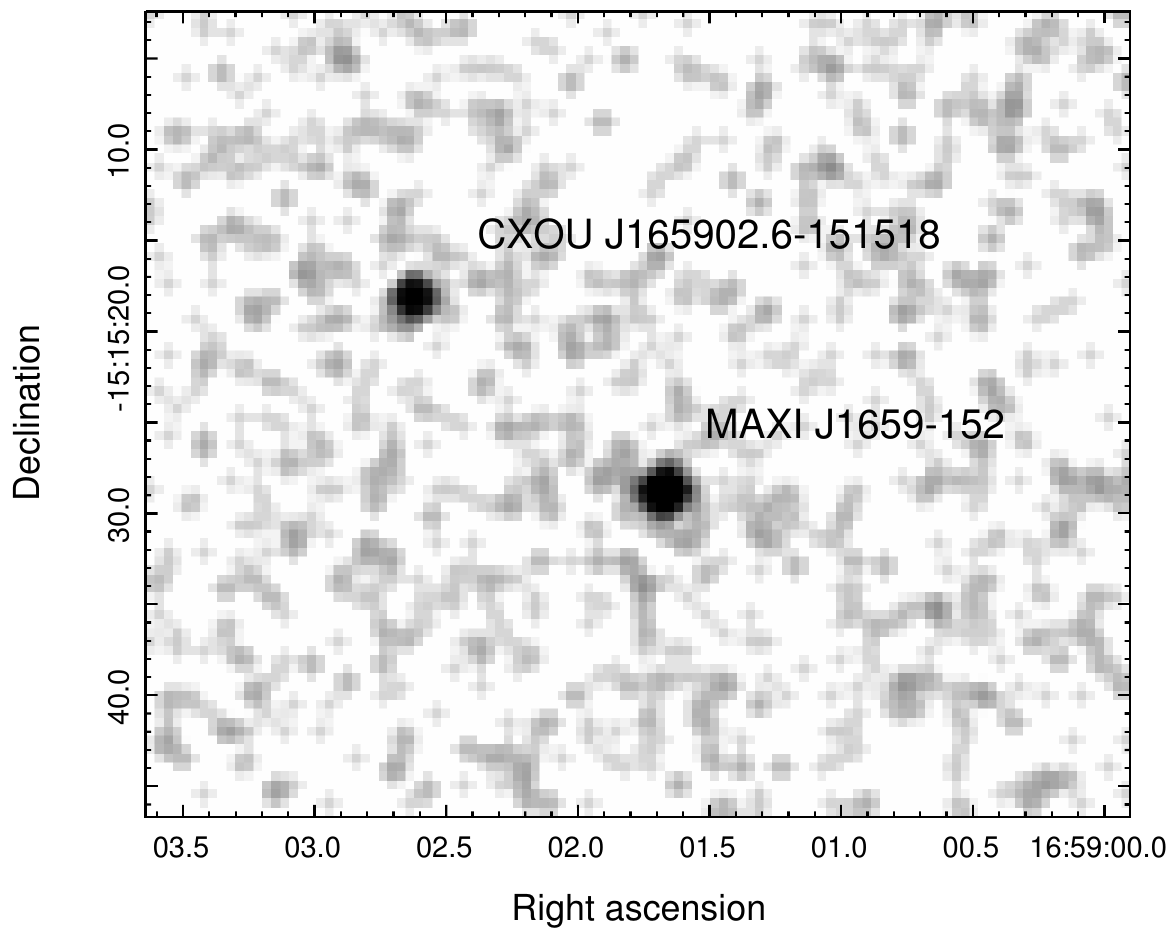}}
\caption{\cha\ 0.5--7 keV image of the area surrounding \maxi. The raw data were smoothed using a Gaussian kernel with a 2-pixel radius. With the exception of the faint uncataloged source CXOU J165902.6--151518, no additional nearby sources were detected. } 
\label{fig:image}
\end{figure}

Source spectra
were extracted from near-circular polygon-shaped regions modeled on the
\cha\ ACIS point-spread function (PSF). The
source extraction regions had a PSF enclosed energy fraction of $\sim$0.97
(for a photon energy of $\sim$1.5 keV) and a radius of $\sim$1\farcs9, except for the extraction region for observation 4, which had an enclosed energy fraction of $\sim$0.98 and a radius of $\sim$2\farcs7 (due to the higher count rate). For the background extraction regions we used annuli centered around the source, with inner radii of $\sim$4\farcs4 (22\arcsec\ for observation 4) and outer radii of 24--25\arcsec\ (45\farcs5 for observation 4). A circular region with a radius of $\sim$4\farcs3 centered around the source to the north-east of \maxi\ (CXOU J165902.6--151518; see Figure \ref{fig:image}) was excluded from the background region. For observation 4 we excluded an additional 10\arcsec$\times$80\arcsec\ rectangle surrounding the read-out streak from the background region. Response files were created using the {\tt mkacisrmf} and {\tt mkarf} tools in CIAO.

The spectra were fitted in the 0.3--7.0 keV range with XSPEC 12.7.1 \citep{ar1996}. Given the low number of source counts per spectrum (as low as 10), we used the C statistic \citep{ca1979}, modified to account for the subtraction of background counts, the so-called W statistic. The spectra were grouped to at least one photon per spectral bin. Following \citet{jomiho2012} all spectra were fitted with an absorbed power law ({\tt tbabs*pegpwrlw} in XSPEC), with the abundances set to {\tt wilm} and the cross sections set to {\tt vern}. Because of high count rates  (see Table \ref{tab:obs}), for observation 4 we also added the pile-up model of \citet{da2001}. 

The $N_{\rm H}$  was first determined from fits to the spectrum of observation 4, which had the highest number of counts, and subsequently it was fixed in all spectral fits. We obtained a value of 0.33(2)$\times10^{22}$ atoms\,cm$^{-2}$, which is somewhat higher than the value of 0.23$\times10^{22}$ atoms\,cm$^{-2}$ used by \citet{jomiho2012} \citep[see also][]{keroma2011}. 

To put to our \cha\ observations in the context of the full outburst decay, we also constructed a 0.3--10 keV light curve from archival \sw/XRT observations, using the online \sw/XRT data products generator\footnote{{\tt http://www.swift.ac.uk/user\_objects/ }} \citep{evbepa2009}. { For  our \cha\ observations,  corresponding}  \sw/XRT 0.3--10 keV count rates { were calculated from our \cha\ spectra, by }simulating \sw/XRT spectra based on the best-fit model parameters and using \sw/XRT response files.

\section{Results}

All the 0.5--7 keV band images were visually inspected for possible features close to \maxi\ that could be related to jet outflows, such as those seen in XTE J1550--564 \citep{cofetz2003}, H 1743--322 \citep{cokafe2005}, and possibly XTE J1752--223 \citep{rajomi2012}. None could be seen. We also created a combined image from 7 of the 8 observations to increase sensitivity; observation 4 was excluded from this because of the prominent read-out streak. The total exposure time for the resulting image, which is shown in Figure \ref{fig:image},  is $\sim$219 ks. Again, no obvious jet-related structures could be identified. 

\begin{figure}[t] 
\centerline{\includegraphics[width=8cm]{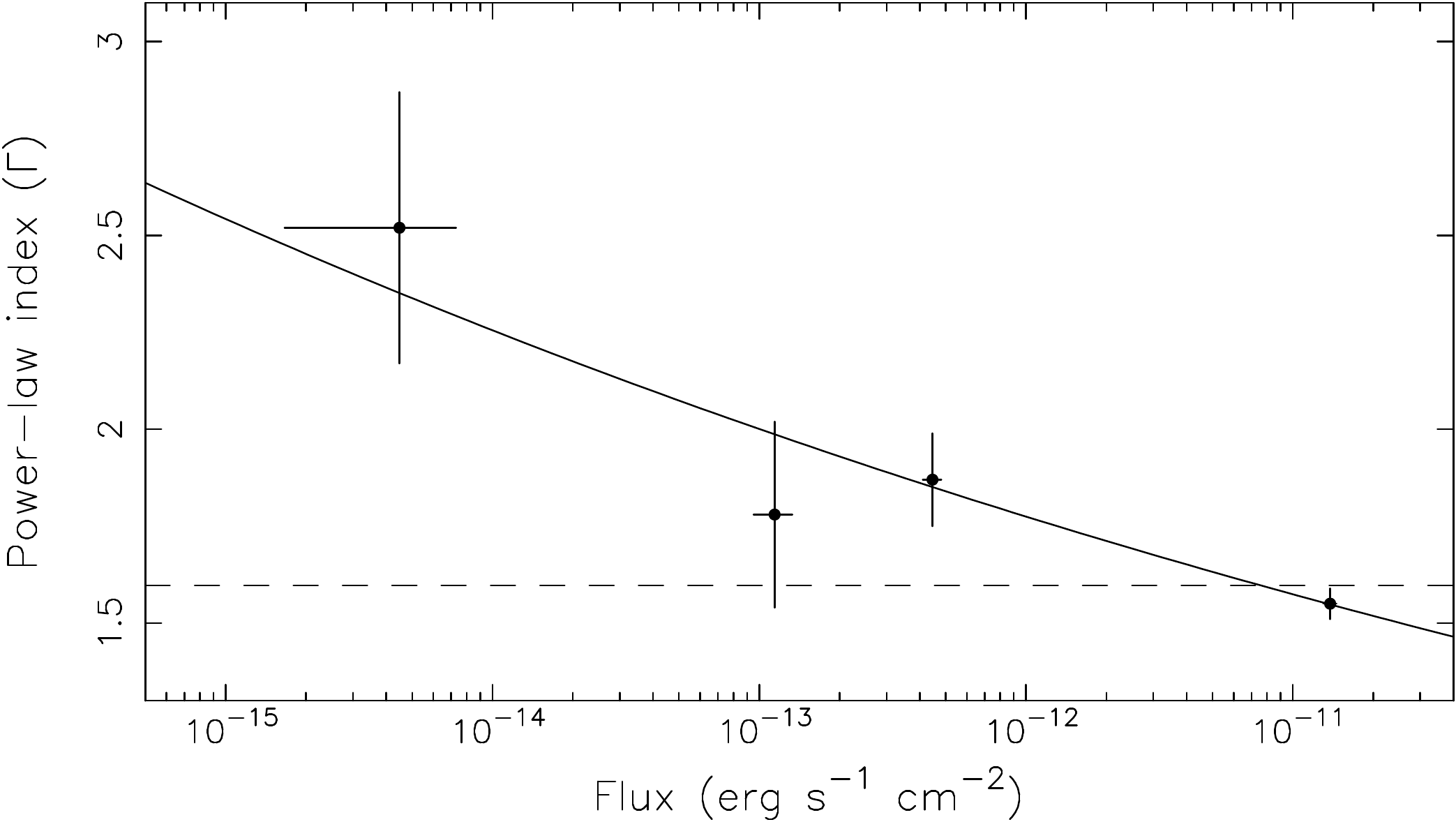}}
\caption{Power-law index of our \cha\ spectra of \maxi\ as a function of the 0.5--10 keV flux. The leftmost data point represents the midpoint of the flux range in observations 3 and 5--8. The horizontal error bar on that data point reflects the observed flux range. The dashed horizontal line shows the best fit with a constant, while the solid line shows the best fit with a power-law.} 
\label{fig:pli}
\end{figure}

Table \ref{tab:obs} lists the results of our spectral fits. Since they had similar count rates, the spectra of observations 3 and 5--8 were fitted simultaneously with their power-law indices tied, as the power-law indices would otherwise be poorly constrained; the normalizations were left free to vary independently. The unabsorbed 0.5--10 keV fluxes measured during the two new \cha\ observations (7 and 8) of \maxi\ are $1.7(9)\times10^{-15}$ erg\,s$^{-1}$\,cm$^{-2}$ and  $3.7(12)\times10^{-15}$ erg\,s$^{-1}$\,cm$^{-2}$. Since observations 7 and 8 were taken only a few days apart, we also made a fit with the normalizations of the power-law component tied, which resulted in a flux  of $2.8(8)\times10^{-15}$ erg\,s$^{-1}$\,cm$^{-2}$.  We find power-law indices between $\sim$1.5 and $\sim$2.5. There appears to be a correlation between the slope of the power law and the flux, as can be seen from Table \ref{tab:obs} and Figure \ref{fig:pli}; the power law steepens as the flux decreases. A fit { to the spectral index versus flux relation in Figure  \ref{fig:pli}} with a constant index is  significantly worse  ($\chi^2$/dof=14.1/3) than one with  a power-law ($\chi^2$/dof=1.00/2). We note that the power-law indices reported here are somewhat higher than the values reported in \citet{jomiho2012}, although they are consistent at the 1$\sigma$ level individually. This is likely the result of our much smaller source extraction regions  ($\sim$2\arcsec instead of 10\arcsec) and the higher $N_{\rm H}$ value that we used in our fits.

In Figure \ref{fig:curve} we show the combined \sw\ and \cha\ light curve of \maxi. The \cha\ data are shown in red. The dashed horizontal line shows the count rate level corresponding to the average flux in the last two \cha\ observations. Around day 200 \maxi\ showed an initial decline towards quiescence. This decline was rapid, with an exponential decay time scale of 5.3$\pm$0.3 days (fitted to the first three \cha\ observations, plus the \sw\ observation near day 200). The best-fit exponential decay is shown as a gray diagonal line. The source had nearly reached the quiescent level at the time of the third \cha\ observation. However, shortly thereafter it showed a nearly 90-day reflare \citep{yawi2011}, during which the flux went up by a factor of $\sim$3000. The second decay was rapid as well; a fit to the two \sw\ data points before the fifth  \cha\ observation yields an exponential decay time scale of 4.8 $\pm$ 0.9 days, hence the $e$-folding times are consistent with being the same.

\section{Discussion}

\begin{figure}[t] 
\centerline{\includegraphics[width=8cm]{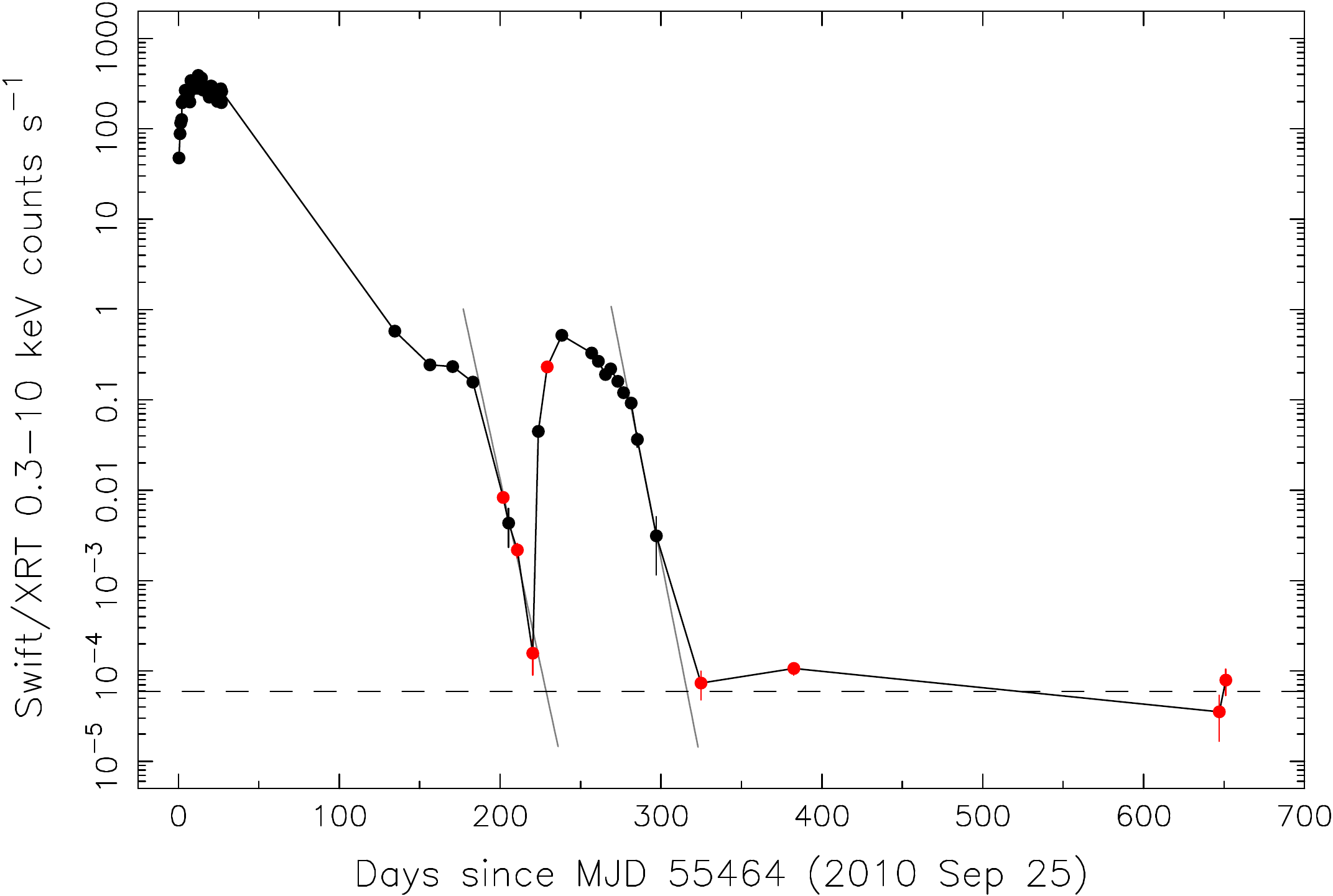}}
\caption{Outburst/quiescence light curve of \maxi. Black data points are \sw/XRT observations; red points show \cha\ observations converted to \sw/XRT count rates. The dashed horizontal line shows the count rate corresponding to the quiescent flux measured in \cha\ observations 7 and 8. The gray diagonal lines shows the best exponential fits to the first and second decay, with $e$-folding times of 5.3 $\pm$ 0.3 and 4.8 $\pm$ 0.9 days, respectively. } 
\label{fig:curve}
\end{figure}

We have presented an analysis of \cha\ observations of  \maxi. Two new observations were analyzed in addition to the earlier six presented in \citet{jomiho2012} { and \citet{plgajo2013}}. These two observations were made more than 320 days after the end of the reflare that was observed at the end of the outburst of \maxi\ (see Figure \ref{fig:curve}). While the flux of observations 7/8 is the lowest value observed in the quiescent state of \maxi,  it is consistent (within 1$\sigma$ errors) with the value measured in observation 5, which was taken close to the end of the reflare. The flux of observation 6 was also within a factor of two of the flux seen in observations 7/8. { The five lowest fluxes seen with \cha\ (obs.\ 3, 5, 6, 7+8) all fall within a factor of $\sim$2.5 of each other. Combined with the fact that these fluxes were measured over a time span of $\sim$430 days, this suggests that this flux range (2.8(8)--7(3)$\times10^{-15}$ erg\,s$^{-1}$\,cm$^{-2}$) represents a relatively stable (within a factor of $\sim$2.5) minimum quiescent flux for \maxi.}

The average flux of observations 7/8 translates into a 0.5--10 keV luminosity of $1.2(3)\times10^{31}$   ($\frac{d}{{\rm 6\,kpc}})^2$ erg\,s$^{-1}$, where the distance $d$ is likely in the range 4.5--8.5 kpc (see Section \ref{sec:intro}). In Figure \ref{fig:orbital} we show the quiescent luminosities of NS (stars) and BH LMXBs (circles) as a function of \porb; \maxi\ is shown as an open circle. As can be seen from this figure there is a clear correlation between the quiescent luminosities of BH LMXBs and their \porb, although there is substantial scatter. The quiescent luminosity of \maxi\ is relatively high compared to some of the systems in the \porb\ range of 4.2--8.3 hr.  This would still be true had we assumed a distance of 4.5 kpc, which would imply a luminosity of 6.8$\times10^{30}$ erg\,s$^{-1}$.  As mentioned in Section \ref{sec:intro}, the mechanism that drives the mass accretion  in quiescence is expected to switch from gravitational-wave losses to evolution of the secondary star around a \porb\ of $\sim$5--10 hr in BH LMXBs, depending on the mass of the secondary \citep{meesna1999}. { As a result, the lowest quiescent BH luminosities are expected to be found in systems with a \porb\ around 5--10 hr.} Based on the data presented in their paper, \citet{gahojo2008} already suggested that such a minimum may exist at a limiting luminosity of a few times 10$^{30}$ \lum. Although the number of systems with \porb\ of a few hours is still very small, the relatively high quiescent luminosity of \maxi, the shortest \porb\ system, may be a further sign of the existence of a minimum quiescent luminosity for BH LMXBs, in a range consistent { with that implied (\porb$\sim$5--10 hr) by the work of \citet{meesna1999}.} 

Additional support for the existence of a minimum quiescent luminosity for BH LMXBs may come from Swift J1357.2--0933, a very faint X-ray transient discovered in early 2011 \citep{krbaba2011}. This system likely contains a BH \citep{arderu2013} and has a short orbital period of 2.8 hr  \citep{cosamu2013}. A {\it Swift}/XRT observation at the end of the outburst, when the source was returning to or had returned to quiescence, yielded  an upper limit on the 0.5--10 keV luminosity of 2$\times10^{31}$ \lum\ \citep{arderu2013}, for a distance of 1.5 kpc \citep{ragrfi2011}. Swift J1357.2--0933 is shown as the gray data point in Figure \ref{fig:orbital}. This upper limit is close to the quiescent luminosity of \maxi, and about a decade higher than one would expect based on extrapolating the general trend seen for BH systems above \porb\ = 4 hr. 
 
Figure \ref{fig:pli} shows a steepening of the spectrum as the flux decreases. This has previously been seen in other sources \citep{cotoka2006,plgajo2013}, but our observations of \maxi\ present one of the clearest examples of spectral softening as a source recedes into quiescence. Our \cha\ observations of \maxi\ allow us to set constraints on the luminosity level at which this softening starts. In observation 4, which showed  the highest luminosity of our observations, the power-law index was found to be 1.55(4), which is consistent with the indices { seen in the hard state at the end of the outburst of \maxi\ \citep{yaalka2012}}. In the second brightest observation (nr.\,1), the index was already significantly higher at 1.87(12). It is therefore likely that in \maxi\ the spectral softening started between 0.5--10 keV fluxes of   $4.5\times10^{-13}$ erg\,s$^{-1}$\,cm$^{-2}$ (obs.\,1) and $1.4\times10^{-11}$ erg\,s$^{-1}$\,cm$^{-2}$ (obs.\,4). This corresponds to fractional Eddington luminosities\footnote{We use an Eddington luminosity  of 1.3$\times10^{38}$ ($M/M_\odot$) \lum.} of (0.18--6.2)$\times10^{-5}$($\frac{d}{{\rm 6\,kpc}})^2$ ($\frac{M}{8\,M_\odot}$). For XTE J1550--564 \citet{cotoka2006} report an average power-law index of 2.25(8) in the 0.5--10 keV flux range of (7--94)$\times10^{-14}$ \flux. For  a distance of 4.38 kpc and a BH mass of 9.1 $M_\odot$ \citep{orstmc2011} this implies that the softening in XTE J1550--564 must have started above a fractional Eddington luminosity of $\sim$1.8$\times10^{-6}$. In H 1743--322 an index of 2.2(6) was measured by  \citet{cotoka2006}  at a 0.5--10 keV flux of 5.0(7)$\times10^{-13}$ \flux. For a distance of 8.5 kpc \citep{stmcre2012} this implies that the softening must have started above a fractional Eddington luminosity of 4.2$\times10^{-6}$ ($\frac{M}{8\,M_\odot}$). {  Combining \cha\ data from ten quiescent BHs (including the ones discusses above), \citet{plgajo2013} find that softening is already ongoing around $3\times10^{-5}$ \ledd\ and plateaus around  $\sim3\times10^{-6}$ \ledd, similar to the range we find for \maxi.}

{ There are also various  reports of softening of X-ray spectra toward quiescence based on {\it RXTE} data \citep[see, e.g.,][]{tocoka2001,wugu2008,dikato2008,rumadu2010,sopado2011}. These authors report that softening already starts at luminosities of $\sim10^{-2}$ \ledd, which  is much higher than the luminosity range implied by \maxi. However, in several of these works \citep{tocoka2001,rumadu2010,sopado2011} the spectra were not corrected for Galactic ridge emission, which provides a natural explanation for the observed softening; the Galactic ridge emission can be fitted with a power law with an index of $\sim$2.14 \citep{re2003}, and an even higher index when an absorbed power law is used. Simulations that we performed  suggest that a ridge contribution of as little as 5\% can already result in detectable softening in a typical {\it RXTE} observation. In other works \citep{wugu2008,dikato2008} attempts were made to correct the spectra for the Galactic ridge emission. We inspected {\it Swift}/XRT archival data of GRO J1655--40 \citep{hokoto2005}, GX 339--4, and H 1743--332 (both this work), taken around the same time as the {\it RXTE} data, or covering the same luminosity range; these data show no evidence for spectral softening occurring around $10^{-2}$ \ledd. Moreover,   at a few times   $10^{-4}$ \ledd\ \citet{wugu2008} find a range of indices (with {\it RXTE}) that is substantially higher ($\sim$1.7--2.5) then the average index found at a few times $10^{-5}$ \ledd\ with \cha\ \citep[$\sim1.7\pm0.1$;][]{plgajo2013}. We therefore suspect that the reports of spectral softening based on  {\it RXTE} data may not be reliable and they should therefore be regarded with some caution.}

As mentioned in Section \ref{sec:intro}, the softening of spectra in quiescence can be explained by a variety of models. Quiescent spectra are generally not of sufficient quality for accurate spectral modeling (beyond a simple power law) and it is therefore difficult to test (and distinguish between) competing  models for quiescent accretion flows. However, the observed softening (and its relation with luminosity) can possibly be used for this purpose. For example, \citet{banaqu2001} showed that for convection-dominated accretion flows  the softening in quiescence is expected to occur at luminosities below $\sim$$10^{-7}$ \ledd, whereas our work shows that softening already starts at luminosities $\sim$20--600 times higher.

The outburst light curve in Figure \ref{fig:curve} shows that \maxi\ had almost reached quiescence around day $\sim$220. However, a major reflare, during which the luminosity increased by a factor of $\sim$3000, occurred soon after. While reflares (or secondary maxima) near the end of an outburst are not uncommon \citep[see, e.g.,][]{chlige1993,tokaka2004,rucumu2012}, the magnitude of this reflare appears to be unusually high. This may be partly due to fact that the secondary maximum is well separated from the main outburst by a brief period of near-quiescence, whereas in other systems it occurred during  (the decay of) the main outburst phase. Given the short orbital period of the system it is possible that the secondary had undergone substantial X-ray heating of its outer layers, possibly resulting in a temporary increase in the mass transfer rate \citep[see, e.g.,][]{aukush1993}. 

Finally, during its outburst \maxi\ showed spectral and timing signatures \citep{kahoal2011} that suggested that the source had crossed the so-called ``jet line'' \citep{fehobe2009} during its transition from the hard state to softer spectral states. Such crossings have been associated with major ejection events, which are observed in the radio, but also on occasion in X-rays. Although the outburst of \maxi\ has been monitored densely in radio, no radio flares were observed around the time at which \maxi\ crossed the jet line \citep{pavabe2013}. Our deep \cha\ images of \maxi\ do not reveal any indications for a major ejection event in X-rays either.

\acknowledgments  J.H.\ would like to thank the members of the Astronomical Institute ``Anton Pannekoek" and SRON Utrecht, where part of this work was done, for their hospitality. This work made use of the UK Swift Science Data Centre at the University of Leicester, NASA's Astrophysics Data System Bibliographic Services, of SAOImage DS9, developed by the Smithsonian Astrophysical Observatory, and of software provided by the Chandra X-ray Center (CXC) in the application packages CIAO. Support for this work was provided by the National Aeronautics and Space Administration through Chandra Award Nos.\ GO1-120498 and GO2-13061X, issued by the Chandra X-ray Observatory Center, which is operated by the Smithsonian Astrophysical Observatory for and on behalf of the National Aeronautics Space Administration under contract NAS8-03060. D.M.R.\ acknowledges support from a Marie Curie Intra-European Fellowship within the 7th European Community Framework Programme under Contract No.\ IEF 274805.


\clearpage
\newpage

\end{document}